# Band Structure Engineering of Interfacial Semiconductors Based on Atomically Thin Lead Iodide Crystals

*Yan Sun, Zhen Huang, Zishu Zhou, Jiangbin Wu, Liujiang Zhou, Yang Cheng, Jinqiu Liu, Chao Zhu, Maotao Yu, Peng Yu, Wei Zhu, Yue Liu, Jian Zhou, Bowen Liu, Hongguang Xie, Yi Cao, Hai Li, Xinran Wang, Kaihui Liu, Xiaoyong Wang, Jianpu Wang, Lin Wang[*], Wei Huang[*]*

Y. Sun, Z. Huang, Z. Zhou, M. Yu, Y. Liu, B. Liu, H. Xie, Y. Cao, Prof. H. Li, Prof. J. Wang, Prof. L. Wang, Prof. W. Huang

Key Laboratory of Flexible Electronics (KLOFE) & Institute of Advanced Materials (IAM), Jiangsu National Synergetic Innovation Center for Advanced Materials (SICAM), Nanjing Tech University (Nanjing Tech), 30 South Puzhu Road, Nanjing 211816, China.

E-mail: iamlwang@njtech.edu.cn; iamwhuang@nwpu.edu.cn

Dr. J. Wu

Ming Hsieh Department of Electrical Engineering, University of Southern California, Los Angeles, CA 90089, USA.

Dr. L. Zhou, Dr. W. Zhu

Center for Nonlinear Studies and Theoretical Division, Los Alamos National Laboratory, Los Alamos, New Mexico 87545, USA.

Y. Cheng, Prof. K. Liu

State Key Laboratory for Mesoscopic Physics, School of Physics Department, Peking University, Beijing 100871, China.

J. Liu, Prof. X. Wang

National Laboratory of Solid State Microstructures, School of Physics, Physics Department, Nanjing University, Nanjing 210093, China.

Dr. C. Zhu,

Center for Programmable Materials, School of Materials Science & Engineering, Nanyang Technological University, Singapore 639798, Singapore.

Dr. P. Yu,

School of Materials Science and Engineering, Sun Yat-sen University, Guangzhou 510275, China.

J. Zhou, Prof. X. Wang,

National Laboratory of Solid-State Microstructures School of Electronic Science and Engineering Collaborative Innovation Center of Advanced Microstructures, Nanjing University, Nanjing 210093, China.

Prof. W. Huang

Shaanxi Institute of Flexible Electronics (SIFE), Northwestern Polytechnical University (NPU), 127 West Youyi Road, Xi'an 710072, Shaanxi, China.

**Abstract:**

To explore new constituents in two-dimensional materials and to combine their best in van der Waals heterostructures, are in great demand as being unique platform to discover new physical phenomena and to design novel functionalities in interface-based devices. Herein, PbI$_2$ crystals as thin as few-layers are first synthesized, particularly through a facile low-temperature solution approach with the crystals of large size, regular shape, different thicknesses and high-yields. As a prototypical demonstration of flexible band engineering of PbI$_2$-based interfacial semiconductors, these PbI$_2$ crystals are subsequently assembled with several transition metal dichalcogenide monolayers. The photoluminescence of MoS$_2$ is strongly enhanced in MoS$_2$/PbI$_2$ stacks, while a dramatic photoluminescence quenching of WS$_2$ and WSe$_2$ is revealed in WS$_2$/PbI$_2$ and WSe$_2$/PbI$_2$ stacks. This is attributed to the effective heterojunction formation between PbI$_2$ and these monolayers, but type I band alignment in MoS$_2$/PbI$_2$ stacks where fast-transferred charge carriers accumulate in MoS$_2$ with high emission efficiency, and type II in WS$_2$/PbI$_2$ and WSe$_2$/PbI$_2$ stacks with separated electrons and holes suitable for light harvesting. Our results demonstrate that MoS$_2$, WS$_2$, WSe$_2$ monolayers with very similar electronic structures themselves, show completely distinct light-matter interactions when interfacing with PbI$_2$, providing unprecedent capabilities to engineer the device performance of two-dimensional heterostructures.

**Introduction**

Heterostructure is a versatile platform to investigate physical phenomena and to build functional devices, owing to the rich physics occurring at the two-dimensional (2D) interface between different materials. The emergence of 2D materials -with strong confinement in one dimension and full freedom in the other two dimensions- have created a truly 2D physics world that reveals many unusual properties and new phenomena. The growing amounts of 2D semiconductors, along with the alterable combination of them in van der Waals (vdWs) heterostructures by means of sophisticated assembly techniques, have offered great flexibility in their band engineering and broad application prospects in optoelectronic devices such as on-chip photodetectors, light emitting diodes, single photon emitter.

As a widely-studied group in 2D materials, transition metal dichalcogenides (TMDs), particularly in the form of monolayers (for instance $MoS_2$, $WS_2$ and $WSe_2$), possess many extraordinary properties such as good photoluminescence (PL) performance and extremely large Coulomb interactions.[1, 2] More importantly, these physical properties can be easily tuned by band alignment, strain effect, doping level and the surrounding environment, which provides extensive possibilities to implement heterostructure engineering based on TMD monolayers.[3-7] The strong p-d orbital hybridization in TMDs near the Fermi energy level affects the band alignment and charge transfer between interfaces, as a result, the PL behavior of TMDs can be tuned.[8-11] While, most TMD/TMD heterostructures exhibit type II band alignment that is favorable to light harvesting,[12] as characterized by the observable interlayer excitons and PL quenching.[13-18] Even though both type-I and type-II band alignment could be formed by interfacing black phosphorus (BP) with TMDs, the PL of TMDs is also quenched because of the narrower bandgap of BP.[19] To date, the enhanced PL emission of TMDs could be achieved with combination of boron nitride (BN), perylene tetracarboxylic dianhydride, ZnO nanorods and metal nanoparticles, as a result of the ultra-smooth and chemical-inert surface of BN, energy transfer between the organic-inorganic interface, strain relaxation and surface plasmon effect, respectively.[20-25] Therefore, it is challenging but desirable to have a general strategy - by means of interfacing one type of 2D semiconductor - to design distinct functionalities on similar types of TMDs via manipulating the charge transfer flow in their heterostructures.

Also, there is always a strong demand for new types of 2D semiconductors that exhibit wide tunability and new optoelectronic properties. As a result, we turn our eyes to $PbI_2$, a layered semiconductor with a large bandgap around 2.5 eV in visible range.[26, 27] $PbI_2$ bulk crystals have good applications in X-ray or γ-ray detection, as well as high-efficient photodetectors and lasers at room temperature.[28-31] Moreover, $PbI_2$ is commonly used as a precursor of lead halide perovskite,[32] a 'star' material for building high-performance photovoltaic devices and light emitting diodes.[33-36] The realization of 2D $PbI_2$ would facilitate us to further design and grow a variety of 2D perovskites, which should show unprecedented potential in ultrathin optoelectronic devices of high compact and high efficiency. Similar to TMDs, $PbI_2$ has hexagonal crystalline pattern with a lead atom layer sandwiched between two layers of iodide atoms, featured by strong covalent in-plane bonding and weak vdWs interactions between layers. [26, 37, 38] The hexagonal lattice structure of layered $PbI_2$ is depicted in **Figure 1**a. The lattice constants ($a = b = 4.559$ Å and $c = 6.990$ Å), are in good conformity with those of TMDs.[26] More importantly, $PbI_2$ manifests itself as a p-type semiconductor, while most TMDs like $MoS_2$ and $WS_2$ prefer to reveal n-type behavior.[39, 40] Thus, the integration of $PbI_2$ with different TMDs can form a rich variety of p–n junctions as the basic building blocks of modern electronic devices. The congruence between $PbI_2$ and TMDs intrigues us to introduce $PbI_2$ into 2D system, and to fabricate TMD/$PbI_2$ heterostructures as a preliminary attempt of band structure engineering based on atomically thin $PbI_2$ crystals.

In this work, the synthesis and characterization of 2D $PbI_2$, and their interfacial interactions with several types of monolayer TMDs are under investigation. We have fabricated atomically thin $PbI_2$ flakes with large size, high crystallinity and hexagonal/triangle shape via a facile solution processing method at atmosphere. As a good example of flexible interface engineering of 2D hybrid system based on PbI$_2$, we have transferred several TMD monolayers (taking MoS$_2$, WS$_2$ and WSe$_2$ for example) on top of PbI$_2$ flakes to assemble TMD/PbI$_2$ heterostructures, and then systematically investigated their optical properties tuned by different interlayer interactions. An enhanced PL emission of MoS$_2$ is observed in MoS$_2$/PbI$_2$ stacks, but conversely, a dramatic PL quenching of WS$_2$ and WSe$_2$ is revealed in WS$_2$/PbI$_2$ and WSe$_2$/PbI$_2$ stacks. As confirmed by first-principles calculations, we attribute these observations to type I band alignment in MoS$_2$/PbI$_2$ stacks, but type II in WS$_2$/PbI$_2$ and WSe$_2$/PbI$_2$ stacks, as corresponding to different charge carrier transfer processes in TMD/PbI$_2$ heterostructures. Our results provoke research effort in PbI$_2$ atomically thin crystals serving as a building block of vdWs heterostructures, which may further enlarge their applications in on-chip optoelectronic devices, such as single photon emitters, light-emitting diodes, photodetectors.

**Results and discussion**

In general, several kinds of methods have been applied to fabricate PbI$_2$ nanosheets, including mechanical exfoliation, physical vapor deposition (PVD) and solution process.[30, 41, 42] In contrast to exfoliation and PVD, solution method has the advantages of high productivity, simple operation, low manufacture cost, and the as-grown crystals are of good crystalline quality, regular shapes and uniform surface.[34, 42] But to date, it still remains a big challenge to synthesize atomically thin PbI$_2$ flakes via solution process. Herein, with the precise control of the concentration of PbI$_2$ precursor solution as well as crystalline nucleation rate, we have successfully obtained PbI$_2$ crystals down to atomic scale as thin as mono- or few- layers. Figure 1b briefly illustrates the preparation process: we first drop cast saturated PbI$_2$ aqueous solution onto a plasma-cleaned Si/SiO$_2$ substrate, and subsequently heat the substrate to 180 °C within 5 minutes to assist PbI$_2$ flakes with nucleation (see more details in Methods section). The whole process is simple and quick, which avoids the rigorous conditions of high temperature and vacuum environment as required by PVD method, and the uncontrollability of the flake thickness, shape and productivity in exfoliation process. The high-yields and good morphology of our grown PbI$_2$ flakes can be revealed by characterizations of optical microscope and atomic force microscopy (AFM). From the optical image in Figure 1c, abundant PbI$_2$ flakes with different thickness are fully deposited on the substrate, exhibiting regular hexagonal or triangle shape with sharp edge. Figure 1e is an AFM scanning image of the selected region marked by white dashed lines in Figure 1c, suggesting the ultra-flat and smooth surface of PbI$_2$ flakes. The height profile indicates that these two typical PbI$_2$ flakes are around 3.0 nm and 4.6 nm thick, respectively. Due to the absorbed water layer and other molecules on the substrates, there might be a small deviation between the real and the AFM-measured thickness of PbI$_2$ flakes, as also can be seen in other 2D materials. [43, 44]

Optical measurements enable us to further check the physical properties of these as-grown PbI$_2$ flakes. Figure 1d shows the Raman spectra of a typical PbI$_2$ flake of 10 nm-thick excited by 532 nm laser (around 2.33 eV). The characteristic peaks around 97 cm$^{-1}$ and 113 cm$^{-1}$ can be determined as A$_{1g}$ and 2LA(M) (overtone of the LA phonon at the M point of 1$^{st}$ Brillouin zone) vibration modes respectively, and the LA(M)+LO(M) (169 cm$^{-1}$) and 2LO(M) (216 cm$^{-1}$) Raman peaks are also visible, in line with the previous reports of PbI$_2$.[41, 45] Furthermore, we could distinguish the subtle split of E$_{2g}$ (75 cm$^{-1}$) vibration peak as an indication of the 4H phase of our PbI$_2$ flakes.[37] Thickness-dependent photoluminescence (PL) of PbI$_2$ flakes were carried out under 290 K and 4 K, as the results shown in Figure 1f and Figure S1, respectively. For PbI$_2$ bulk crystals, a strong emission peak around 512 nm is observed. Owing to quantum confinement, the position of PL emission peak shows a continuous blue-shift as the thickness decreases, accompanied with the gradually dropped intensity. When it comes to few-layers, the PL signal becomes too weak to recognize at room temperature, because of the direct-indirect semiconductor transition occurred around 3 layers. [45, 46] However, the PL emission of PbI$_2$ as thin as few layers at cryogenic temperatures is still very remarkable with three distinct peaks, as corresponding to excitonic emission, traps states in bulk and traps related to surface quality. [47]

On basis of atomically thin PbI$_2$ crystals with high-yields and good quality, we further interface MoS$_2$ monolayers with these PbI$_2$ flakes. The thickness of the chosen PbI$_2$ flakes for MoS$_2$/PbI$_2$ stacks are within the range of around 3-9 nm (although we could synthesize even thinner flakes), which are more than 4 layers and less than 12 layers, as they can keep 2D features and possess a direct bandgap simultaneously.[45, 46] And for the same reason, MoS$_2$ monolayers are selected instead of thicker ones.[1] The exfoliated MoS$_2$ monolayers are used to cap the as-grown PbI$_2$ flakes by dry transfer method to construct vdWs heterostructures, as schematically shown in **Figure 2**a (see Methods). The side view of MoS$_2$/PbI$_2$ stacks is depicted by a cartoon image in Figure 2b. Figure 2d presents an optical microscope image of one heterostructure sample, in which a large continuous MoS$_2$ piece covers several PbI$_2$ flakes. The contour of the PbI$_2$ flakes are outlined by black dashed lines, and two of them are completely encapsulated by the monolayer MoS$_2$. The thickness of the larger one (about 8 μm$^2$) is around 3.0 nm, and the other (about 3 μm$^2$) is around 12.4 nm. Raman spectroscopy has been extensively used to study the effects of doping, strain and interlayer interactions in TMDs. [3, 4, 48, 49] As for the Raman spectrum of MoS$_2$/PbI$_2$ stacks shown in Figure 2c, the Raman peaks around 96 cm$^{-1}$ and 113 cm$^{-1}$ assigned to PbI$_2$ are still observable. When comparing with the Raman spectrum of the monolayer MoS$_2$ alone, in the MoS$_2$/PbI$_2$ stacks the A$_{1g}$ peak has a blue-shift ~ 0.5 cm$^{-1}$ and the E$^1_{2g}$ peak has a red-shift ~ 1 cm$^{-1}$. It may have many origins such as vdWs interactions, strain, doping and laser-induced thermal effect. [23, 49-52] A very possible scenario for the change of the Raman peak positions is the strong interlayer interaction between MoS$_2$ and PbI$_2$, as evidenced by the appearance of layer breathing mode in the ultra-low frequency Raman spectra (denoted in Figure 2c). It is known that the layer breathing mode could only be seen in 2D heterostructures with good-quality interface. [54] Figure 2e illustrates the PL emission mapping under 488 nm laser excitation with integration from 1.79 to 1.88 eV of the entire heterostructure. Strong PL signals are detected in the monolayer $MoS_2$ region, but even more enhanced PL emission in the regions of $MoS_2/PbI_2$ stacks, as can be seen from that the brilliant yellow profile of $PbI_2$ flakes under $MoS_2$ is quite clear.

**Figure 3**a carefully compares the PL spectra of the sample shown in Figure 2, with the signal taken from the $MoS_2$ alone and $MoS_2/PbI_2$ region, respectively. When forming $MoS_2/PbI_2$ stacks, the PL signal assigned to monolayer $MoS_2$ shifts to high energy position, and the intensity increases by several folds. We have fabricated more than 10 samples, all of which show similar optical behavior with the increasing amplitude in the range of 160%-600% if calculated by $\frac{PL(TMD/PbI_2)}{PL(TMD)}$. Furthermore, similar PL behavior is also observed in $PbI_2/MoS_2$ stacks in which $PbI_2$ is on top of $MoS_2$ (see Figure S3). To better understand the origin of these changes, we fit the PL spectra with three Lorentzian peaks, as corresponding to negative trion ($X^-$) at ~1.82 eV, neutral exciton (X) at ~1.86 eV and B exciton at ~1.98 eV.[4, 5] From the analysis, the proportion of the trion ($X^-$) is greater than that of the neutral exciton (X) in monolayer $MoS_2$ alone, because of the n-type trait of $MoS_2$.[40] With the addition of $PbI_2$, the PL enhancement and blue-shift of $MoS_2$ emission peak is because the neutral exciton (X) emission significantly increases, while the trion ($X^-$) emission keeps almost unchanged. This clearly reveals that the doping effect is not the main origin of the PL enhancement of $MoS_2$ in $MoS_2/PbI_2$ heterostructures, very differently from other 2D systems with high emission of $MoS_2$. For instance, in $MoS_2/BN$ heterostructures, the underlying BN flakes prevent the n-type doping effect of $MoS_2$ from $SiO_2$ substrates effectively- which causes the reduction of trion emission ($X^-$) and then

relative enhancement of neutral exciton emission (X)- leading to the stronger PL signal of $MoS_2$ (the detailed comparison can be seen in Figure S4).[54, 55] In $MoS_2/PbI_2$ heterostructures the inrush of external electrons and holes rather than doping itself, accounts for the enhanced emission of the neutral exciton (X). In order to figure out the origin of these charge carries, we have carried out density functional theory (DFT) calculations on the constructed heterostructures (see more details in Methods section and Figure S5). As shown in Figure 3b, $MoS_2/PbI_2$ heterostructures show type-I band alignment owing to the large band gap of $PbI_2$ and strong built-in potentials. The excitons originating from the absorption of photons in $PbI_2$ partly separate under the effect of build-in potentials, and quickly accumulate in $MoS_2$ by charge transfer process. These exotic carries occupy the neutral exciton (X) energy level of $MoS_2$ and release energy by radiative recombination ultimately, leading to the improvement and blueshift of PL emission of $MoS_2$. It is also noted that in $MoS_2/PbI_2$ stacks, the B exciton emission peak drops and moves to higher energy level. Take the existence of heavy Pb atoms in $PbI_2$ and the interlayer coupling between $MoS_2$ and $PbI_2$ into consideration, the change of B exciton emission might be caused by strong spin-orbital coupling. Furthermore, we find that the PL behavior of $MoS_2$ on top of two $PbI_2$ flakes with different thicknesses are almost the same (see Figure S7), and thus we think that the optical interference effect of $PbI_2$ with different thicknesses is subtle. Time-resolved photoluminescence (TRPL) is highly suitable to analyze and determine the fast charge carrier dynamics in semiconductors. In Figure 3c of TRPL spectra, the PL lifetime of $MoS_2$ in $MoS_2/PbI_2$ stacks (~75.9 ps) have little change compared with that of $MoS_2$ alone (~66.8 ps), as $MoS_2$ plays an acceptor role in the heterostructure. The almost coincident rising edge indicates that the charge transfer process between $PbI_2$ and $MoS_2$ is too fast (less than 25ps) to distinguish in our experimental setup. Unfortunately, we are unable to obtain any information about the PL lifetime of $PbI_2$ flakes, as the signal is too weak to detect.

As being typical members of TMDs, WS$_2$ and WSe$_2$ exhibit many similar behaviors to MoS$_2$. Following this rationale, we further use monolayer WS$_2$ and WSe$_2$ instead of MoS$_2$ to build WS$_2$/PbI$_2$ and WSe$_2$/PbI$_2$ stacks. The Raman spectra of these stacks excited by 488 nm laser are shown in **Figure 4**a. The emergence of the Raman peaks characteristic of WS$_2$ and WSe$_2$ also occurs in heterostructures, indicate the high-quality of the prepared stacks. Because of the additional absorption by the PbI$_2$, the peak position of A$_{1g}$ mode exhibits a red shift of ~1.4 cm$^{-1}$, and the increasing resonant peak of 2LA(M) mode surpasses the one of E$_{2g}^1$ mode in the Raman spectra of WS$_2$/PbI$_2$ stacks. In the Raman spectra of WSe$_2$/PbI$_2$ stacks, the peak positions of E$_{2g}^1$ and A$_{1g}$ modes keep the same with the case of isolated WSe$_2$, and the emergence of a new resonant Raman peak (LA(M)+ZA(M)) verifies the strong interaction between WSe$_2$ and PbI$_2$. In contrast to the PL enhancement in MoS$_2$/PbI$_2$ stacks, the PL emission of WS(Se)$_2$ in WS(Se)$_2$/PbI$_2$ stacks decreases (WS$_2$, 5%-40%; WSe$_2$, 4%-30% in all the samples we measured) with a red-shift by about 10 meV (see Figure 4b,c). The optical image and PL mapping in Figure 4d,e intuitively shows this PL quenching effect, where the WSe$_2$/PbI$_2$ region is dramatically dark compared to the bright zone of monolayer WSe$_2$ alone. Our DFT calculations (see Figure 4f) reveal that both WS$_2$/PbI$_2$ and WSe$_2$/PbI$_2$ heterostructures exhibit type-II band alignment, so that the conduction band of the interface is originated from PbI$_2$ side and the valence band from WS$_2$ or WSe$_2$ side. The different flow directions of electrons and holes lead to the loss of neutral excitons, as a result, the emission of trions with low-energy position becomes dominant. Therefore, the total decrease and redshift of WS(Se)$_2$ emission peak in WS(Se)$_2$/PbI$_2$ stacks is observed. The opposite effect of PbI$_2$ on MoS$_2$ and WS(Se)$_2$ also implies that, the different types of band alignment (instead of doping effect), are the dominant factor of determining and manipulating the optical properties of TMD/PbI$_2$ interfacial semiconductors. Therefore, PbI$_2$ atomically thin crystals show great potential in the band structure engineering as a new building block of 2D heterostructures.

**Conclusion**

In summary, atomically thin PbI$_2$ flakes of high quality have been successfully synthesized for the first time. A series of atomic-scale heterostructures constructed by PbI$_2$ and TMD monolayers - namely MoS$_2$, WS$_2$, WSe$_2$/PbI$_2$ - are under investigation to illustrate the unprecedent possibilities and capabilities of 2D PbI$_2$ in the band structure engineering. The optical measurement results show that the addition of PbI$_2$ brings distinctive impacts on different TMDs of similar electronic structure, originating from different types of band alignment. Concretely speaking, an enhanced PL of MoS$_2$ results from the high emission efficiency of accumulated charges in the form of type-I band alignment in MoS$_2$/PbI$_2$ stacks, but the separated electron-hole pairs lead to the dramatically quenched PL effect of WS$_2$ or WSe$_2$ in type-II WS$_2$/PbI$_2$ and WSe$_2$/PbI$_2$ stacks. Atomic-level PbI$_2$-based heterostructures, provide us more freedom to deterministically manipulate the semiconducting interface properties such as the exciton behavior and surface energy transfer process, by selecting the appropriate constituents in the vast family of 2D materials.

**Experimental Section**

*Fabrication*: PbI$_2$ precursor solution was prepared by dissolving PbI$_2$ powder (Sigma-Aldrich) in DI water (1mg/mL), and heated with stirring at 90 °C until PbI$_2$ powder was dissolved completely. Then the solution was drop-cast on an oxygen plasma-cleaned SiO$_2$/Si substrate at room temperature, followed by heating the substrate to 180 °C within 5 minutes to help PbI$_2$ nanosheets nucleate. The monolayer TMDs are obtained by mechanical exfoliation method and then transferred onto the pre-grown PbI$_2$ flakes by PPC-assisted dry transfer technique.

*Characterization*: The topography and height profile were measured by AFM (Park XE7) in non-contact mode. Micro-photoluminescence (PL) spectra of PbI$_2$ flakes were measured by 410 nm solid-state laser excitation source with the laser beam being focused to a spot size of ~2 μm, and PL spectra of pristine TMDs and TMDs/PbI$_2$ stacks were excited by 488 nm continuous wave laser with the laser beam being focused to a spot size of ~500 nm. The PL signals were recorded with a monochromator and a liquid-nitrogen-cooled charge-coupled device (CCD). The Raman spectra were acquired using a micro-Raman system (WITec alpha 300R) equipped with 1800 grooves per millimeter gratings and a liquid-nitrogen-cooled CCD detector. For time-resolved photoluminescence (TRPL) measurements, the excitation pulse laser (410 nm beam, 100 fs, 80 MHz) was focused by a microscopic objective (100×; NA = 0.95) onto the sample at normal incidence. The backscattered signal filtered by a proper long-pass was collected using time-correlated single-photon counting, which has a resolution of ~25 ps. All the measurements are conducted under room temperature, unless stated otherwise.

*Computational details*: The density functional theory (DFT) calculations were carried out by using the Vienna *ab initio* Simulation Package (VASP). [56] The exchange correlation interaction is treated within the generalized gradient approximation (GGA) parameterized by the Perdew, Burke, and Ernzerhof (PBE). [57] Electronic wave functions were built in the plane wave basis sets with a kinetic energy cutoff of 400 eV. All the atoms in models were fully optimized using the optimized Becke88 van der Waals (optB88-vdW) functional until the force on each atom is less than 0.01 eV/Å. [58] The reciprocal space was sampled with a k-grid density of $0.02 \times 2\pi$Å$^{-1}$ for the structure optimization and $0.01 \times 2\pi$Å$^{-1}$ for the calculations of electronic structures. In our models, a vacuum layer of 20 Å was used to isolate neighboring periodic images in all systems. Four-layer PbI$_2$ was utilized to mimic the experimental PbI$_2$ flake. The thicker PbI$_2$ flake was not considered here since our test indicates that it has little effect on the band alignments of heterojunctions. It is important to mention that although

DFT underestimates the electronic band gap, it provides good agreement with the optical band gap in these low-dimensional systems. [59]

**Supporting Information**
Supporting Information is available from the Wiley Online Library or from the author.

**Acknowledgements**
This work was supported by the National Natural Science Foundation of China (61801210, 91733302, 11474164, 61634001, 11574147), Natural Science Foundation of Jiangsu Province (BK20180686, BK20150043, BK20150064), the Joint Research Program between China and European Union (2016YFE0112000), the National Basic Research Program of China-Fundamental Studies of Perovskite Solar Cells (2015CB932200), the National Science Fund for Distinguished Young Scholars (61725502), the Synergetic Innovation Center for Organic Electronics and Information Displays. L.W. gratefully acknowledges support from the Chinese Thousand Talents Plan for Young Professionals. L.Z. and W.Z. acknowledge support from the U.S. Department of Energy through the LANL/LDRD Program and the Center for Nonlinear Studies for this work.

Received: ((will be filled in by the editorial staff))
Revised: ((will be filled in by the editorial staff))
Published online: ((will be filled in by the editorial staff))

**Conflict of interest**
The authors declare no competing financial interests.

**References**

[1] A. Splendiani, L. Sun, Y. Zhang, T. Li, J. Kim, C. Y. Chim, G. Galli, F. Wang, *Nano Lett.* **2010**, *10*, 1271.

[2] K. F. Mak, C. Lee, J. Hone, J. Shan, T. F. Heinz, *Phys. Rev. Lett.* **2010**, *105*, 136805.

[3] H. J. Conley, B. Wang, J. I. Ziegler, R. F. Haglund, Jr., S. T. Pantelides, K. I. Bolotin, *Nano Lett.* **2013**, *13*, 3626.

[4] S. Mouri, Y. Miyauchi, K. Matsuda, *Nano Lett.* **2013**, *13*, 5944.

[5] M. S. Kim, C. Seo, H. Kim, J. Lee, D. H. Luong, J. H. Park, G. H. Han, J. Kim, *ACS Nano* **2016**, *10*, 6211.

[6] J. S. Ross, S. Wu, H. Yu, N. J. Ghimire, A. M. Jones, G. Aivazian, J. Yan, D. G. Mandrus, D. Xiao, W. Yao, X. Xu, *Nat. Commun.* **2013**, *4*, 1474.

[7] K. F. Mak, K. He, C. Lee, G. H. Lee, J. Hone, T. F. Heinz, J. Shan, *Nat. Mater.* **2012**, *12*, 207.

[8] K. Zhao, Z. Deng, X. C. Wang, W. Han, J. L. Zhu, X. Li, Q. Q. Liu, R. C. Yu, T. Goko, B. Frandsen, L. Liu, F. Ning, Y. J. Uemura, H. Dabkowska, G. M. Luke, H. Luetkens, E. Morenzoni, S. R. Dunsiger, A. Senyshyn, P. Boni, C. Q. Jin, *Nat. Commun.* **2013**, *4*, 1442.

[9] G. Zhao, C. Lin, Z. Deng, G. Gu, S. Yu, X. Wang, Z. Gong, Y. J. Uemera, Y. Li, C. Jin, *Sci. Rep.* **2017**, *7*, 14473.

[10] N. R. Wilson, P. V. Nguyen, K. Seyler, P. Rivera, A. J. Marsden, Z. P. Laker, G. C. Constantinescu, V. Kandyba, A. Barinov, N. D. Hine, *Sci. Adv.* **2017**, *3*, e1601832.

[11] G.-B. Liu, W.-Y. Shan, Y. Yao, W. Yao, D. Xiao, *Phys. Rev. B* **2013**, *88*, 085433.

[12] B. Zheng, C. Ma, D. Li, J. Lan, Z. Zhang, X. Sun, W. Zheng, T. Yang, C. Zhu, G. Ouyang, G. Xu, X. Zhu, X. Wang, A. Pan, *J. Am. Chem. Soc.* **2018**, *140*, 11193.

[13] X. Hong, J. Kim, S. F. Shi, Y. Zhang, C. Jin, Y. Sun, S. Tongay, J. Wu, Y. Zhang, F. Wang, *Nat. Nanotechnol.* **2014**, *9*, 682.

[14] S. Tongay, W. Fan, J. Kang, J. Park, U. Koldemir, J. Suh, D. S. Narang, K. Liu, J. Ji, J. Li, *Nano Lett.* **2014**, *14*, 3185.

[15] M. H. Chiu, C. Zhang, H. W. Shiu, C. P. Chuu, C. H. Chen, C. Y. Chang, C. H. Chen, M. Y. Chou, C. K. Shih, L. J. Li, *Nat. Commun.* **2015**, *6*, 7666.

[16] P. Rivera, J. R. Schaibley, A. M. Jones, J. S. Ross, S. Wu, G. Aivazian, P. Klement, K. Seyler, G.

Clark, N. J. Ghimire, J. Yan, D. G. Mandrus, W. Yao, X. Xu, *Nat. Commun.* **2015**, *6*, 6242.

[17] J. Zhang, J. Wang, P. Chen, Y. Sun, S. Wu, Z. Jia, X. Lu, H. Yu, W. Chen, J. Zhu, G. Xie, R. Yang, D. Shi, X. Xu, J. Xiang, K. Liu, G. Zhang, *Adv. Mater.* **2016**, *28*, 1950.

[18] M. Baranowski, A. Surrente, L. Klopotowski, J. M. Urban, N. Zhang, D. K. Maude, K. Wiwatowski, S. Mackowski, Y. C. Kung, D. Dumcenco, A. Kis, P. Plochocka, *Nano Lett.* **2017**, *17*, 6360.

[19] J. Yuan, S. Najmaei, Z. Zhang, J. Zhang, S. Lei, P. M. Ajayan, B. I. Yakobson, J. Lou, *ACS Nano* **2015**, *9*, 555.

[20] A. Boulesbaa, V. E. Babicheva, K. Wang, I. I. Kravchenko, M.-W. Lin, M. Mahjouri-Samani, C. B. Jacobs, A. A. Puretzky, K. Xiao, I. Ivanov, C. M. Rouleau, D. B. Geohegan, *ACS Photonics* **2016**, *3*, 2389.

[21] Y. Li, J. D. Cain, E. D. Hanson, A. A. Murthy, S. Hao, F. Shi, Q. Li, C. Wolverton, X. Chen, V. P. Dravid, *Nano Lett.* **2016**, *16*, 7696.

[22] M. R. Habib, H. Li, Y. Kong, T. Liang, S. M. Obaidulla, S. Xie, S. Wang, X. Ma, H. Su, M. Xu, *Nanoscale* **2018**, *10*, 16107.

[23] S. Wang, X. Wang, J. H. Warner, *ACS Nano* **2015**, *9*, 5246.

[24] H. Jeong, H. M. Oh, A. Gokarna, H. Kim, S. J. Yun, G. H. Han, M. S. Jeong, Y. H. Lee, G. Lerondel, *Adv. Mater.* **2017**, *29*, 1700308.

[25] Z. Q. Wu, J. L. Yang, N. K. Manjunath, Y. J. Zhang, S. R. Feng, Y. H. Lu, J. H. Wu, W. W. Zhao, C. Y. Qiu, J. F. Li, S. S. Lin, *Adv. Mater.* **2018**, *30*, 1706527.

[26] I. C. Schlüter, M. Schlüter, *Phys. Rev. B* **1974**, *9*, 1652.

[27] A. S. Toulouse, B. P. Isaacoff, G. Shi, M. Matuchová, E. Kioupakis, R. Merlin, *Phys. Rev. B* **2015**,

*91*, 165308.

[28] R. Street, S. Ready, K. Van Schuylenbergh, J. Ho, J. Boyce, P. Nylen, K. Shah, L. Melekhov, H. Hermon, *J. Appl. Phys.* **2002**, *91*, 3345.

[29] X. Liu, S. T. Ha, Q. Zhang, M. de la Mata, C. Magen, J. Arbiol, T. C. Sum, Q. Xiong, *ACS Nano* **2015**, *9*, 687.

[30] M. Zhong, L. Huang, H.-X. Deng, X. Wang, B. Li, Z. Wei, J. Li, *J. Mater. Chem. C* **2016**, *4*, 6492.

[31] J. Zhang, T. Song, Z. Zhang, K. Ding, F. Huang, B. Sun, *J. Mater. Chem. C* **2015**, *3*, 4402.

[32] S. Chen, G. Shi, *Adv. Mater.* **2017**, *29*, 1605448.

[33] N. Wang, L. Cheng, R. Ge, S. Zhang, Y. Miao, W. Zou, C. Yi, Y. Sun, Y. Cao, R. Yang, Y. Wei, Q. Guo, Y. Ke, M. Yu, Y. Jin, Y. Liu, Q. Ding, D. Di, L. Yang, G. Xing, H. Tian, C. Jin, F. Gao, R. H. Friend, J. Wang, W. Huang, *Nat. Photonics* **2016**, *10*, 699.

[34] J. Liu, Y. Xue, Z. Wang, Z. Q. Xu, C. Zheng, B. Weber, J. Song, Y. Wang, Y. Lu, Y. Zhang, Q. Bao, *ACS Nano* **2016**, *10*, 3536.

[35] L. Niu, X. Liu, C. Cong, C. Wu, D. Wu, T. R. Chang, H. Wang, Q. Zeng, J. Zhou, X. Wang, W. Fu, P. Yu, Q. Fu, S. Najmaei, Z. Zhang, B. I. Yakobson, B. K. Tay, W. Zhou, H. T. Jeng, H. Lin, T. C. Sum, C. Jin, H. He, T. Yu, Z. Liu, *Adv. Mater.* **2015**, *27*, 7800.

[36] Y. Cao, N. Wang, H. Tian, J. Guo, Y. Wei, H. Chen, Y. Miao, W. Zou, K. Pan, Y. He, H. Cao, Y. Ke, M. Xu, Y. Wang, M. Yang, K. Du, Z. Fu, D. Kong, D. Dai, Y. Jin, G. Li, H. Li, Q. Peng, J. Wang, W. Huang, *Nature* **2018**, *562*, 249.

[37] R. Zallen, M. L. Slade, *Solid State Commun.* **1975**, *17*, 6.

[38] M. Ando, M. Yazaki, I. Katayama, H. Ichida, S. Wakaiki, Y. Kanematsu, J. Takeda, *Phys. Rev. B* **2012**, *86*, 155206.

[39] R. Street, S. Ready, F. Lemmi, K. Shah, P. Bennett, Y. Dmitriyev, *J. Appl. Phys.* **1999**, *86*, 2660.

[40] B. Liu, W. Zhao, Z. Ding, I. Verzhbitskiy, L. Li, J. Lu, J. Chen, G. Eda, K. P. Loh, *Adv. Mater.* **2016**, *28*, 6457.

[41] P. Wangyang, H. Sun, X. Zhu, D. Yang, X. Gao, *Mater. Lett.* **2016**, *168*, 68.

[42] W. Zheng, Z. Zhang, R. Lin, K. Xu, J. He, F. Huang, *Adv. Electron. Mater.* **2016**, *2*, 1600291.

[43] H. Li, Z. Yin, Q. He, H. Li, X. Huang, G. Lu, D. W. Fam, A. I. Tok, Q. Zhang, H. Zhang, *Small* **2012**, *8*, 63.

[44] X. Lu, M. I. B. Utama, J. Lin, X. Luo, Y. Zhao, J. Zhang, S. T. Pantelides, W. Zhou, S. Y. Quek, Q. Xiong, *Adv. Mater.* **2015**, *27*, 4502.

[45] M. Yagmurcukardes, F. M. Peeters, H. Sahin, *Phys. Rev. B* **2018**, *98*, 085431.

[46] C. Cong, J. Shang, L. Niu, L. Wu, Y. Chen, C. Zou, S. Feng, Z.-J. Qiu, L. Hu, P. Tian, Z. Liu, T. Yu, R. Liu, *Adv. Opt. Mater.* **2017**, *5*, 1700609.

[47] I. Baltog, M. Baibarac, S. Lefrant, *J. Phys.: Condens. Matter* **2009**, *21*, 025507.

[48] M. H. Chiu, M. Y. Li, W. Zhang, W. T. Hsu, W. H. Chang, M. Terrones, H. Terrones, L. J. Li, *ACS Nano* **2014**, *8*, 9649.

[49] X. Zhang, X. F. Qiao, W. Shi, J. B. Wu, D. S. Jiang, P. H. Tan, *Chem. Soc. Rev.* **2015**, *44*, 2757.

[50] H. Li, J. B. Wu, F. Ran, M. L. Lin, X. L. Liu, Y. Zhao, X. Lu, Q. Xiong, J. Zhang, W. Huang, H. Zhang, P. H. Tan, *ACS Nano* **2017**, *11*, 11714.

[51] K. G. Zhou, F. Withers, Y. Cao, S. Hu, G. Yu, C. Casiraghi, *ACS Nano* **2014**, *8*, 9914.

[52] M. Buscema, G. A. Steele, H. S. J. van der Zant, A. Castellanos-Gomez, *Nano Res.* **2015**, *7*, 561.

[53] S. Sahoo, A. P. S. Gaur, M. Ahmadi, M. J. F. Guinel, R. S. Katiyar, *The Journal of Physical Chemistry C* **2013**, *117*, 9042.

[54] L. Liang, J. Zhang, B. G. Sumpter, Q. H. Tan, P. H. Tan, V. Meunier, *ACS Nano* **2017**, *11*, 11777.

[55] M. D. Tran, J. H. Kim, H. Kim, M. H. Doan, D. L. Duong, Y. H. Lee, *ACS Appl. Mater. Interfaces* **2018**, *10*, 10580.

[56] G. Kresse, J. Furthmüller, *Phys. Rev. B* **1996**, *54*, 11169.

[57] J. P. Perdew, K. Burke, M. Ernzerhof, *Phys. Rev. Lett.* **1996**, *77*, 3865.

[58] A. D. Becke, *Phys. Rev. A* **1988**, *38*, 3098.

[59] Y. Gong, J. Lin, X. Wang, G. Shi, S. Lei, Z. Lin, X. Zou, G. Ye, R. Vajtai, B. I. Yakobson, H. Terrones, M. Terrones, B. K. Tay, J. Lou, S. T. Pantelides, Z. Liu, W. Zhou, P. M. Ajayan, *Nat. Mater.* **2014**, *13*, 1135.

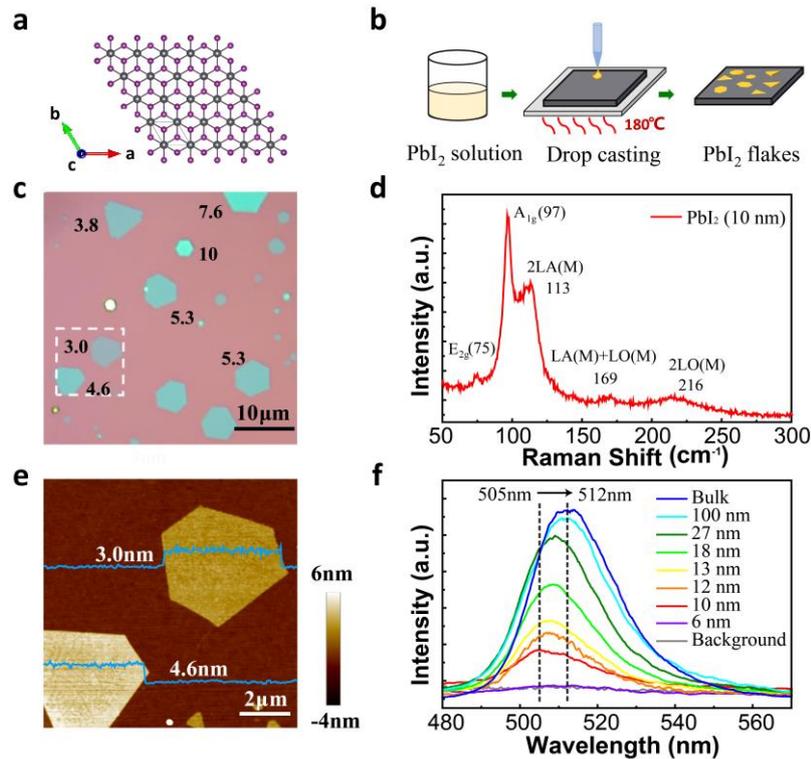

**Figure 1.** a) The hexagonal lattice of PbI$_2$ from top view. The dark grey balls represent lead (Pb) atoms and the purple balls represent iodine (I) atoms. b) Schematic illustration of synthesizing atomically thin PbI$_2$ flakes via solution method. PbI$_2$ aqueous solution was prepared and drop-cast onto Si/SiO$_2$ substrate, followed by heating the substrate to 180 °C within 5 minutes to assist PbI$_2$ flakes to nucleate. c) The optical image of the as-grown PbI$_2$ flakes with regular hexagonal or triangle shape with sharp edge. The labelled numbers represent the thickness of the corresponding flakes. Scale bar: 10 μm. d) The Raman spectra of a typical 10 nm-thick PbI$_2$ flake excited by 532 nm laser. e) Atomic force microscopy image of the selected region outlined by white dashed lines in (c), confirms the smooth and flat surface of PbI$_2$ flakes. The height profile of the two flakes indicates that their thickness are around 3.0 nm and 4.6 nm, respectively. Scale bar: 2 μm. f) The photoluminescence spectra of PbI$_2$ flakes with different thicknesses measured at room temperature. With decreasing the crystal thickness, the position of emission peak has a continuous blue-shift and the intensity decreases correspondingly.

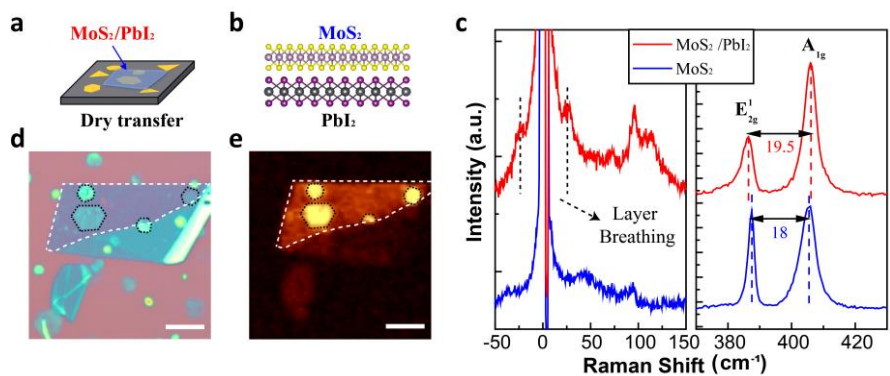

**Figure 2.** a) Schematic image of MoS$_2$/PbI$_2$ stacks. The exfoliated MoS$_2$ monolayer is used to cap the as-grown PbI$_2$ flakes by dry transfer method. b) The lattice structure of MoS$_2$/PbI$_2$ stacks from side view. The lilac balls represent molybdenum (Mo) atoms and the yellow balls represent sulphur (S) atoms. c) The Raman spectra of pristine MoS$_2$ monolayer and MoS$_2$/PbI$_2$ stacks. d) Optical image of MoS$_2$/PbI$_2$ stacks with a large monolayer MoS$_2$ piece capping several PbI$_2$ flakes. The hexagonal PbI$_2$ flakes are marked by black dashed lines, and two of them are completely encapsulated by the monolayer MoS$_2$ whose outlines are denoted by white dashed lines. Scale bar: 5 μm. e) The photoluminescence mapping image of the same heterostructure shown in (d), spectrally integrated from 1.79 to 1.88 eV, in which the brilliant yellow denotes the high photoluminescence emission intensity. Scale bar: 5 μm.

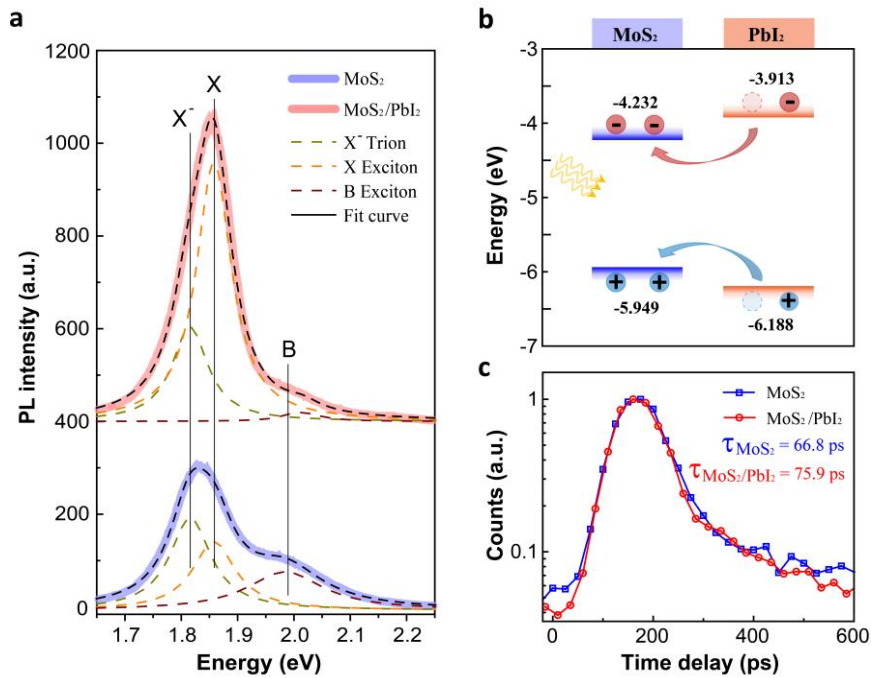

**Figure 3**. a) The photoluminescence spectra of pristine $MoS_2$ monolayer and $MoS_2/PbI_2$ stacks with 488 nm excitation, in the analysis of three Lorentzian peaks corresponding to negative trion ($X^-$) at ~1.82 eV, neutral exciton (X) at ~1.86 eV and B exciton at ~1.98 eV. In $MoS_2/PbI_2$ stacks, the position assigned to $MoS_2$ shifts to high energy, and the total intensity increases by several folds with the $X^-$ emission unchanged. b) Type-I band alignment for $MoS_2/PbI_2$ stacks with illustratively showing the transfer of photoexcited carriers within the heterostructure, as calculated by density functional theory. The inrush of electrons and holes from $PbI_2$ to $MoS_2$ make the contribution from neutral exciton emission (X) increase. c) The time-resolved photoluminescence spectra of pristine $MoS_2$ monolayer and $MoS_2/PbI_2$ stacks, showing that the photoluminescence lifetime of $MoS_2$ in $MoS_2/PbI_2$ stacks (~75.9 ps) is almost the same with that of pristine $MoS_2$ (~66.8 ps).

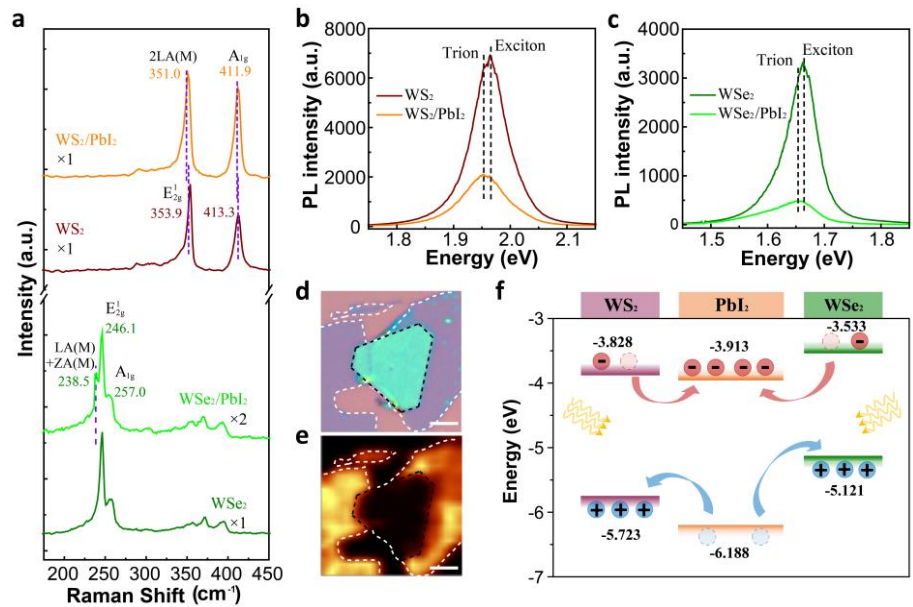

**Figure 4.** a) The Raman spectra of pristine WS(Se)$_2$ monolayer and WS(Se)$_2$/PbI$_2$ stacks. The red shift of A$_{1g}$ mode and the increasing resonant peak of 2LA(M) mode indicate the strong interaction between WS$_2$ and PbI$_2$ layers. The emergence of a new resonant Raman peak (LA(M)+ZA(M)) also verifies the strong interaction between WSe$_2$ and PbI$_2$ layers. b-c) The photoluminescence spectra of b) pristine WS$_2$ monolayer and WS$_2$/PbI$_2$ stacks, c) pristine WSe$_2$ monolayer and WSe$_2$/PbI$_2$ stacks. In WS(Se)$_2$/PbI$_2$ stacks, the photoluminescence of WS(Se)$_2$ show a significant quenching, as well as a red-shift in peak position. d) Optical image of WSe$_2$/PbI$_2$ stacks with monolayer WSe$_2$ capping a big PbI$_2$ flake. The contour of the hexagonal PbI$_2$ flake is marked by black dashed lines, and the monolayer MoS$_2$ white dashed lines. Scale bar: 3 μm. e) The photoluminescence mapping image of the same heterostructure shown in (d), spectrally integrated from 1.77 to 1.94 eV, in which the photoluminescence of WSe$_2$ is quenched on location of PbI$_2$. Scale bar: 3 μm. f) Type-II band alignment for WS$_2$/PbI$_2$ and WSe$_2$/PbI$_2$ stacks, with the illustrative transfer of photoexcited carriers within the heterostructures as calculated by density functional theory. The conduction band minimum of the interface is originated from PbI$_2$ and the valence band maximum from WS(Se)$_2$. The separation of excitons leads to the photoluminescence quenching of WS(Se)$_2$ in WS(Se)$_2$/PbI$_2$ stacks.

**The table of contents entry**

**PbI$_2$** with unique electronic structure, can be synthesized down to atomic scale by solution method, and executed to construct versatile interfacial semiconductors via band alignment engineering. As an illustrative example, the photoluminescence of MoS$_2$ is enhanced due to the type-I nature of MoS$_2$/PbI$_2$ stacks, while the dramatically quenching of WS$_2$ and WSe$_2$ occurs in type-II WS$_2$/PbI$_2$ and WSe$_2$/PbI$_2$ stacks.

**Keywords**: band engineering

Yan Sun, Zhen Huang, Zishu Zhou, Jiangbin Wu, Liujiang Zhou, Yang Cheng, Jinqiu Liu, Chao Zhu, Maotao Yu, Peng Yu, Wei Zhu, Yue Liu, Jian Zhou, Bowen Liu, Hongguang Xie, Yi Cao, Hai Li, Xinran Wang, Kaihui Liu, Xiaoyong Wang, Jianpu Wang, Lin Wang[*], Wei Huang[*]

**Band Structure Engineering of Interfacial Semiconductors Based on Atomically Thin Lead Iodide Crystals**

ToC Figure

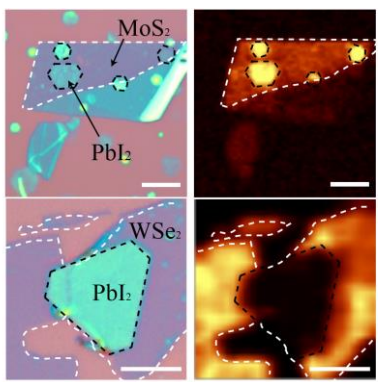

Supporting Information:

**Band Structure Engineering of Interfacial Semiconductors Based on Atomically Thin Lead Iodide Crystals**

*Yan Sun, Zhen Huang, Zishu Zhou, Jiangbin Wu, Liujiang Zhou, Yang Cheng, Jinqiu Liu, Chao Zhu, Maotao Yu, Peng Yu, Wei Zhu, Yue Liu, Jian Zhou, Bowen Liu, Hongguang Xie, Yi Cao, Hai Li, Xinran Wang, Kaihui Liu, Xiaoyong Wang, Jianpu Wang, Lin Wang[*], Wei Huang[*]*

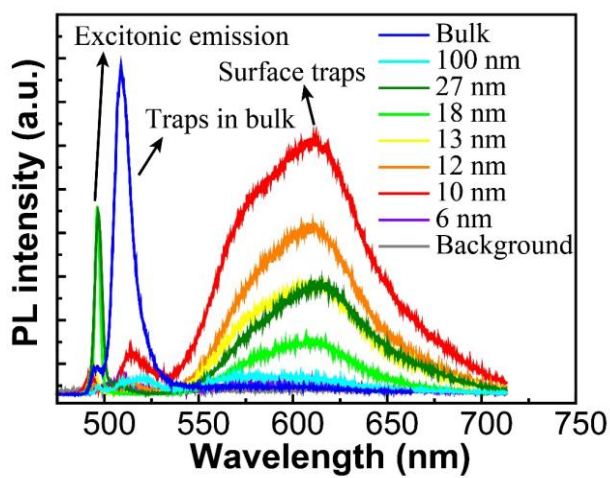

**Figure S1**. The photoluminescence spectra of layered PbI$_2$ with different thickness excited by 410 nm laser at 4 K, with three distinct peaks corresponding to excitonic emission, traps states in bulk and traps related to surface quality.

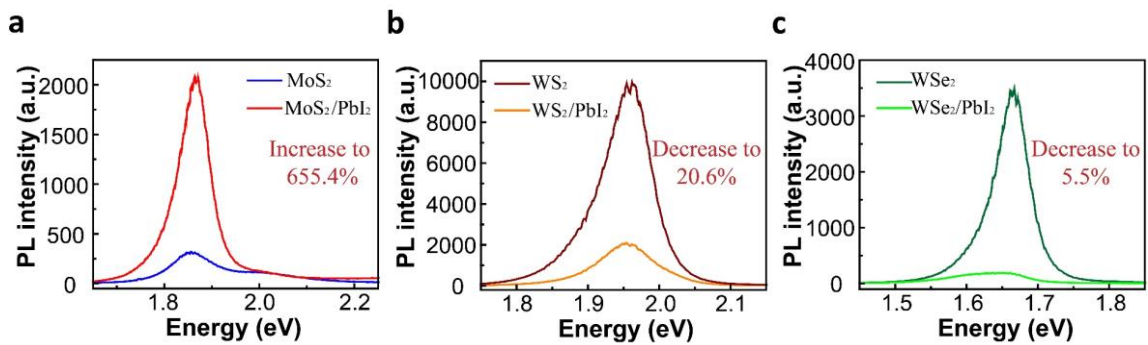

**Figure S2**. The photoluminescence behavior shown in other MoS$_2$/PbI$_2$, WS$_2$/PbI$_2$ and WSe$_2$/PbI$_2$ stack samples.

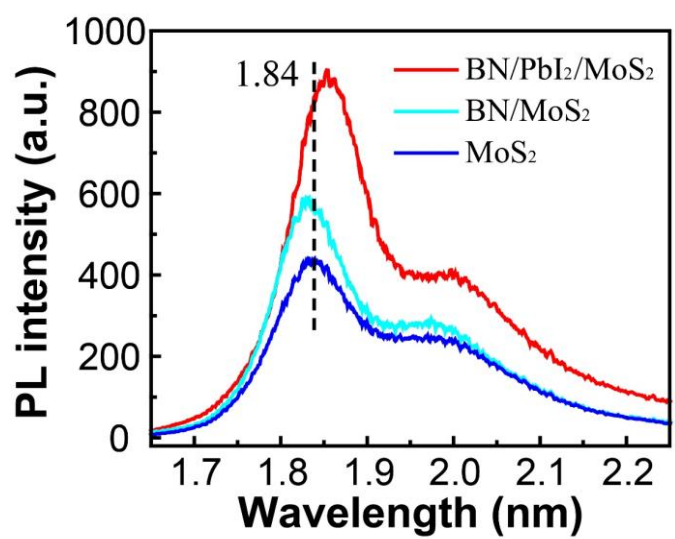

**Figure S3**. The PL properties of BN/PbI$_2$/MoS$_2$ heterostructure on SiO$_2$ substrate (top: BN; middle: PbI$_2$; bottom: MoS$_2$) in which MoS$_2$ is always interfaced with SiO$_2$. Since the hygroscopic nature of PbI$_2$, an ultrathin BN (~2.4 nm) is used to protect PbI$_2$ from the air and all the transfer process was performed in glove-box. In the region of MoS$_2$ covered by PbI$_2$, the PL intensity still increases significantly and the peak position have a blue-shift compared to that of MoS$_2$ alone and MoS$_2$ covered by BN, similar to what observed in MoS$_2$/PbI$_2$ as shown in the main text.

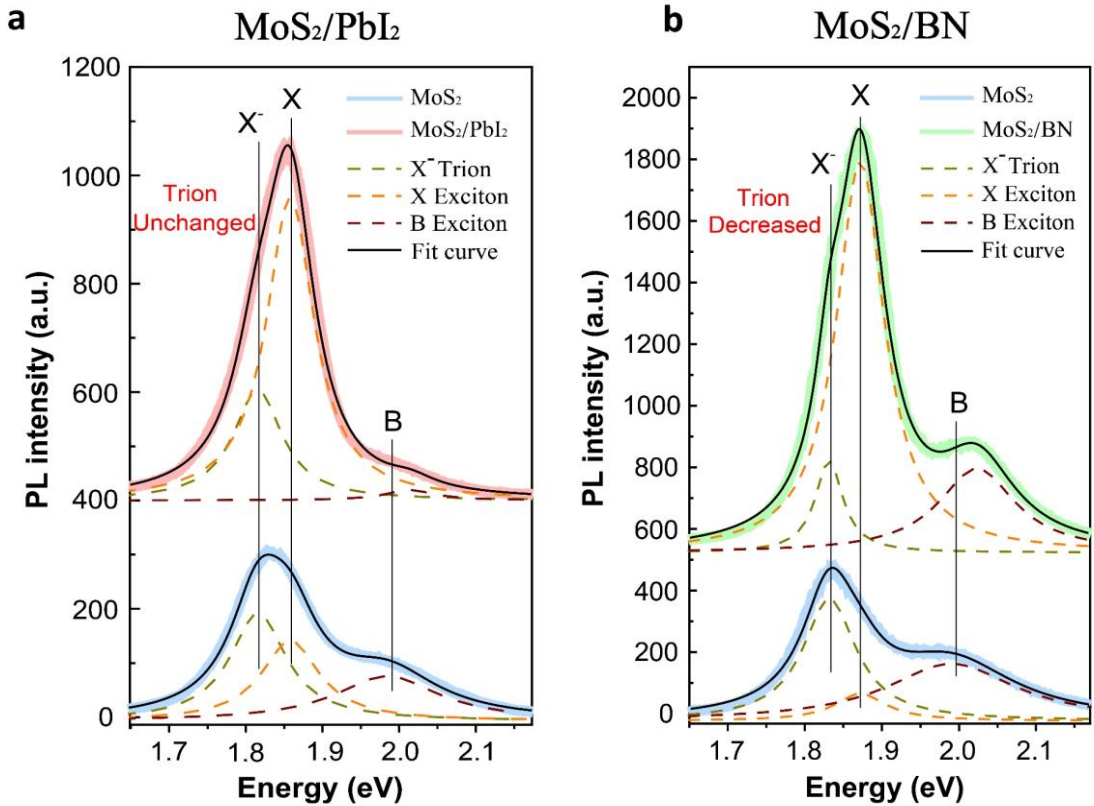

**Figure S4**. The contrastive PL spectra with Lorentz fitting of individual MoS$_2$ monolayers, MoS$_2$/PbI$_2$ and MoS$_2$/BN heterostructures. Compared with individual MoS$_2$ monolayers, the spectral weight of negative trions in MoS$_2$ is almost unchanged (or slightly increased) in MoS$_2$/PbI$_2$ heterostructures, while decreases significantly in MoS$_2$/BN heterostructures. The underlying BN effectively prevent the n-doping effect of SiO$_2$ substrates on MoS$_2$ monolayers, and the charge density effect is the major origin of PL enhancement in MoS$_2$/BN heterostructures.[1] However, the PL enhancement of MoS$_2$ on the location of PbI$_2$ results from the significantly increasing excitons in the type I form of band alignment.

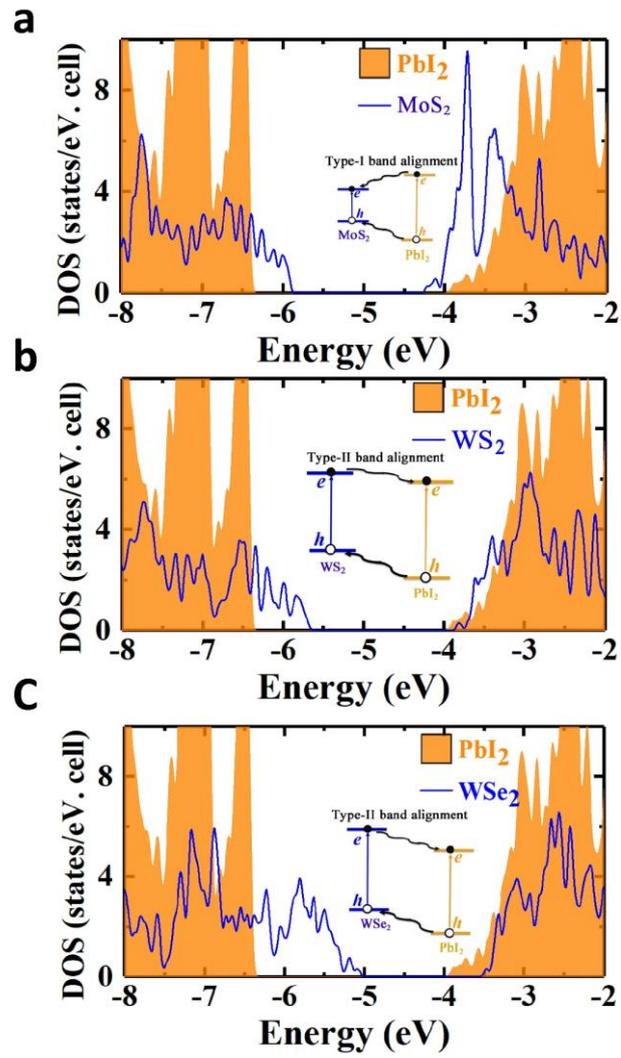

**Figure S5**. Partial density of states of a) $MoS_2/PbI_2$, b) $WS_2/PbI_2$ and c) $WSe_2/PbI_2$ stacks.

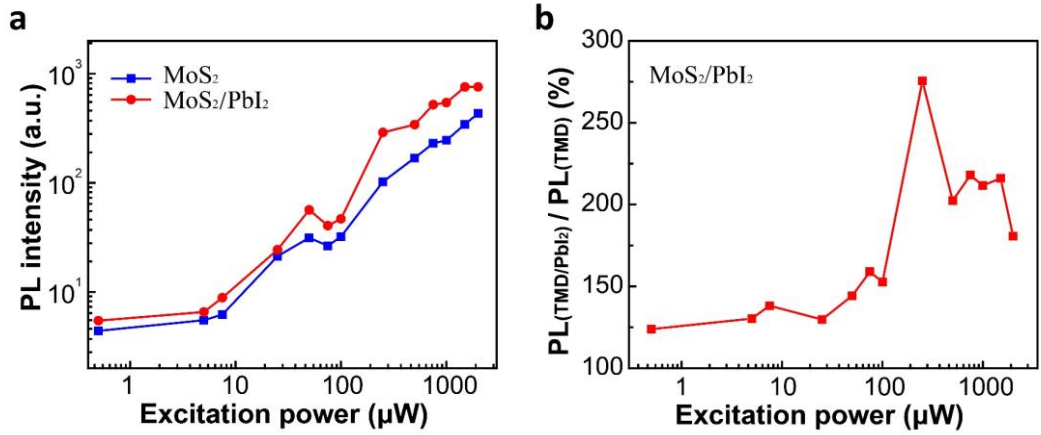

**Figure S6.** a) The relationship between the photoluminescence (PL) intensity of the emission peak and the excitation power for $MoS_2$ alone and $MoS_2/PbI_2$ stack. b) The amplitude of PL enhancement ratio as a function of excitation power. All is in logarithmic coordinate. The enhancement ratio is around ~140% under low excitation, and increases to 220% when the excitation power is high enough (>250 μW). This transition implies that the PL enhancement is limited by traps under low excitation, which is consistent with our inference that the substantial enhancement behavior is caused by charge transfer from $PbI_2$. The amplitude of PL enhancement is calculated by $\frac{PL(TMD/PbI_2)}{PL(MoS_2)}$, where $PL_{(TMD/PbI_2)}$ represents the PL intensity of $MoS_2$ emission peak in $MoS_2/PbI_2$ stacks, and $PL_{TMD}$ in $MoS_2$ alone.

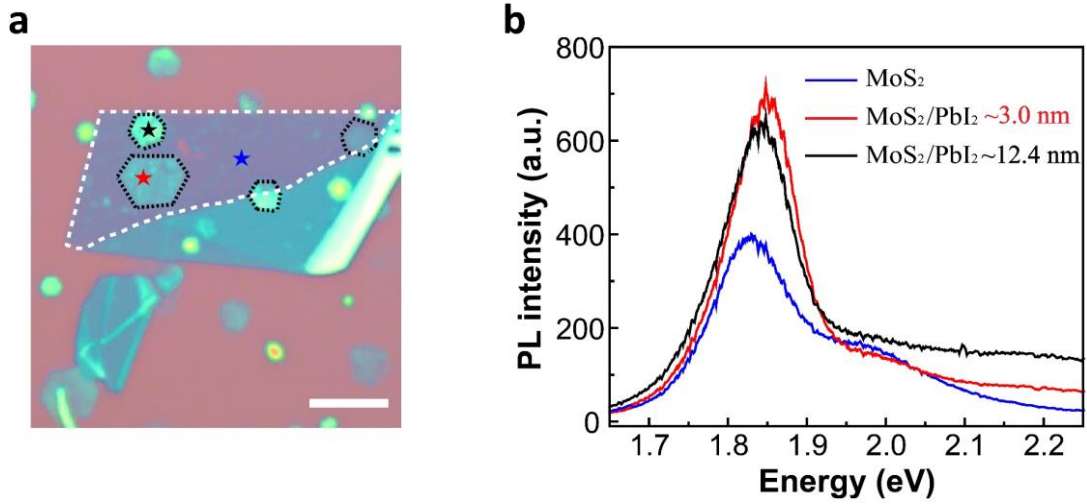

**Figure S7.** a) Optical image of the MoS$_2$/PbI$_2$ stacks shown in the Figure 2 in the main text, in which the contours of MoS$_2$ and PbI$_2$ flakes are marked by while and black lines, and the star symbols denote the locations where the PL spectra are taken. Scale bar: 5 μm. b) The PL spectra of MoS$_2$ alone and MoS$_2$/PbI$_2$ stacks with PbI$_2$ flakes being around 3.0 nm and 12.4 nm thick, respectively. The PL enhancement behavior of MoS$_2$ monolayer in different MoS$_2$/PbI$_2$ stacks is not sensitive to the thickness of PbI$_2$, which demonstrates that the optical interference effect of PbI$_2$ with different thicknesses is subtle.

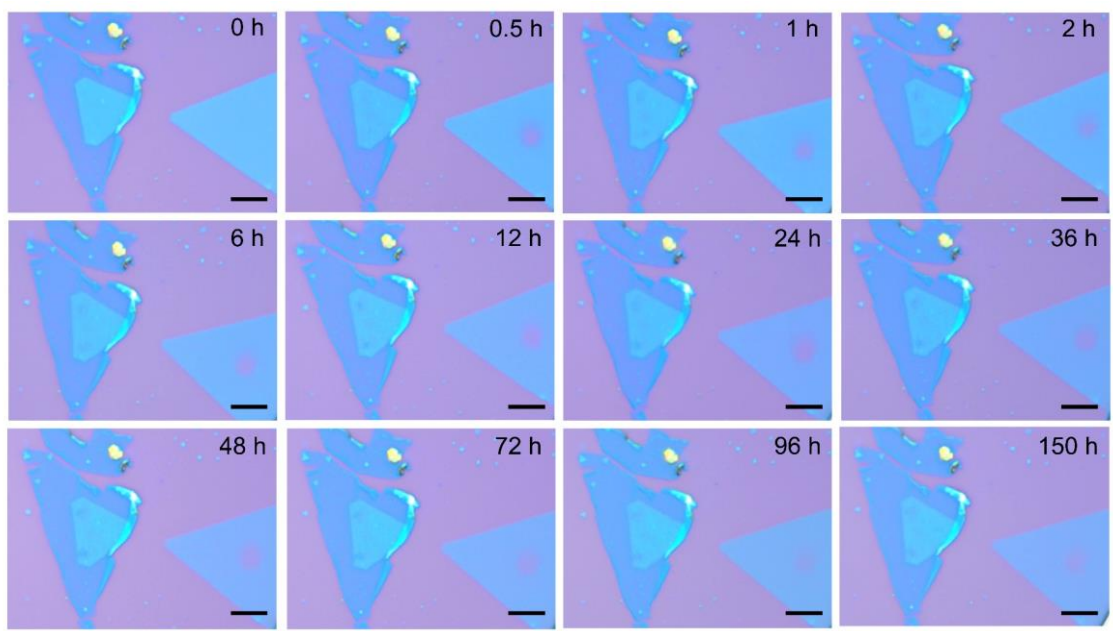

**Figure S8**. The optical images of two PbI$_2$ flakes with similar thickness (one of which is covered by a MoS$_2$ monolayer and the other is bared), when exposed to atmosphere for 0, 0.5, 1, 2, 6, 12, 24, 36, 48, 72, 96, 150 hours, respectively. Because of the intrinsic hygroscopic nature, PbI$_2$ flakes is not very stable in air, while the degradation process is prevented efficiently when capped with MoS$_2$. The three holes on the heterostructure and one hole on the bare PbI$_2$ flake are caused by the high-power laser radiation. Scale bar: 5μm.

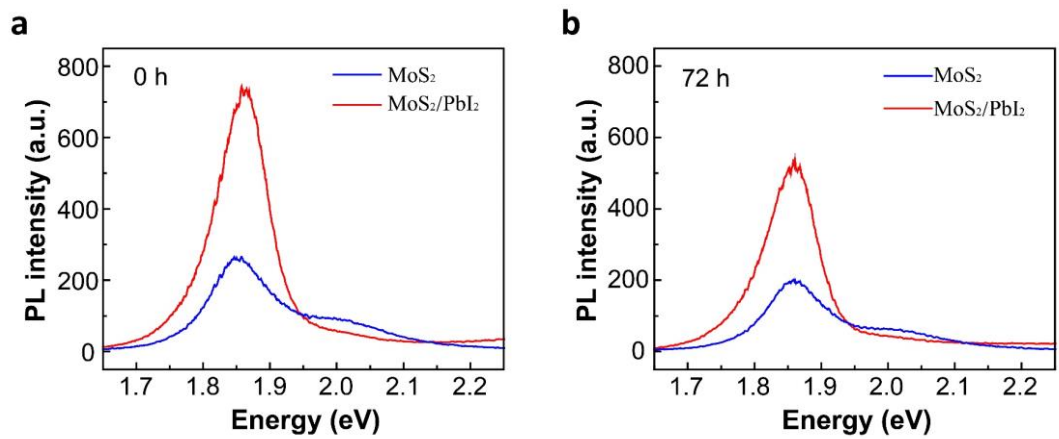

**Figure S9.** The PL spectra of $MoS_2$ alone and $MoS_2/PbI_2$ stacks a) in fresh conditions and b) after stored in air for 72 hours, revealing that the PL enhancement ratio changes from ~280% to ~260%.

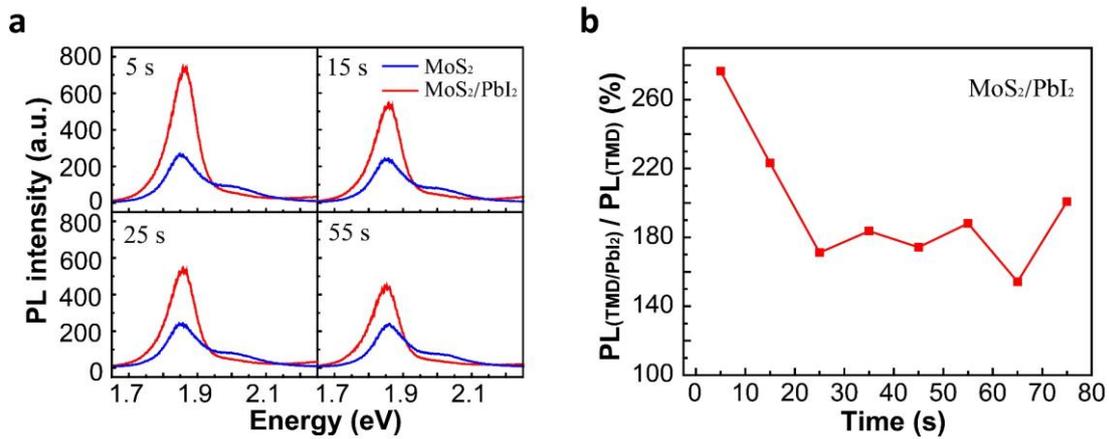

**Figure S10**. The stability of MoS$_2$/PbI$_2$ heterostructure under laser exposure. a) The PL spectra after the sample exposed to laser for 5, 15, 25, 55 seconds. b) The relation between the amplitude of PL enhancement change and exposed time. The laser used here is Continuous Wave 488 nm laser and the excitation power is 2 mW μm$^{-2}$.

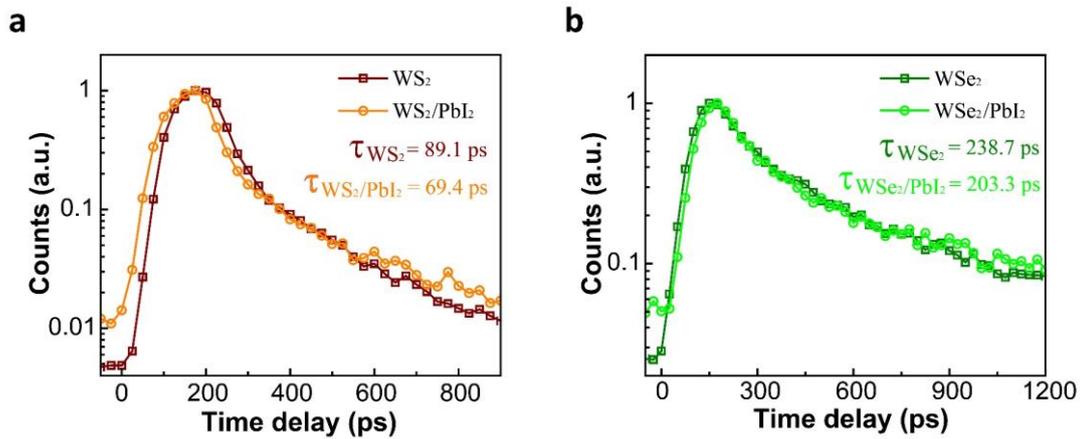

**Figure S11**. Time-resolved photoluminescence spectra of a) WS$_2$ alone and WS$_2$/PbI$_2$ stacks, b) WSe$_2$ alone and WSe$_2$/PbI$_2$ stacks. The PL lifetime of WS(Se)$_2$ in WS(Se)$_2$/PbI$_2$ stacks is slightly shorter than WS(Se)$_2$ alone. As WS(Se)$_2$/PbI$_2$ are type-II heterostructures, the photo-induced electrons in WS(Se)$_2$ will partly transfer to PbI$_2$ side, and this extra charge loss channel in WS(Se)$_2$ results in shorter lifetime. It should be noted that the PL efficiency of TMD monolayer is generally low due to the defect effect,[2, 3] therefore the trap state is still the dominant factor in energy loss process compared to the impact of charge transfer, and then the lifetime change of TMDs in heterostructures is not obvious.

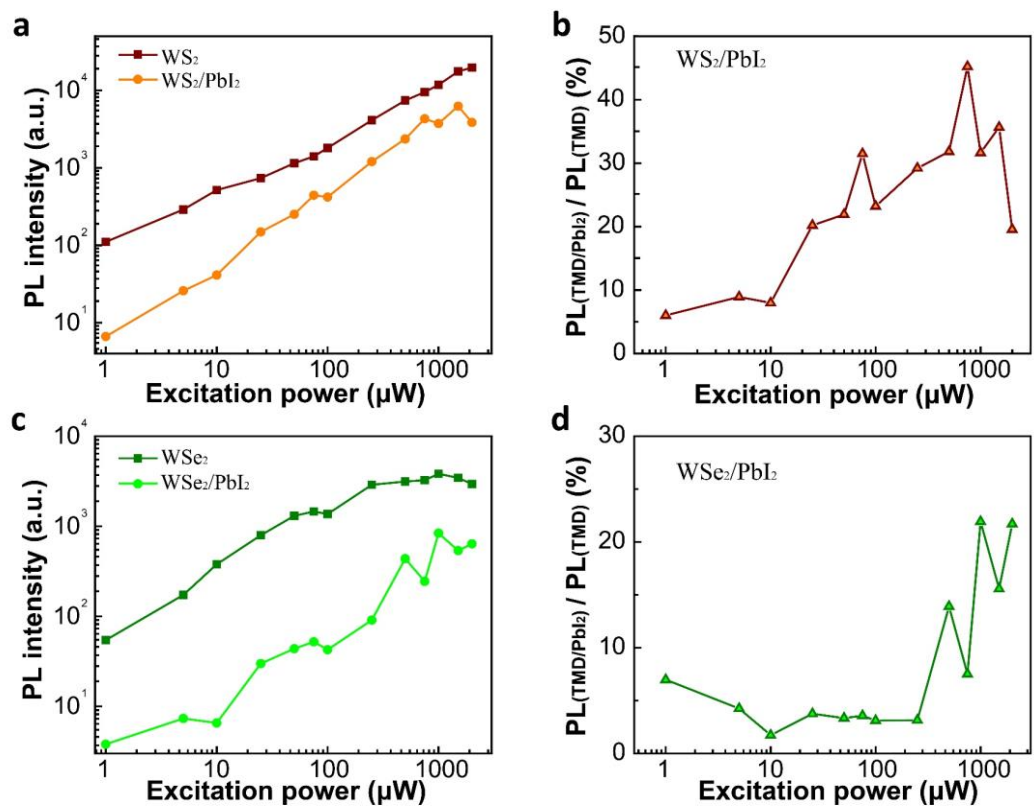

**Figure S12.** The relationship of the photoluminescence intensity and the decreasing amplitude of WS$_2$ or WSe$_2$ as a function of the excitation power for a,b) pristine WS$_2$ and WS$_2$/PbI$_2$ stacks, c,d) WSe$_2$ and WSe$_2$/PbI$_2$ stacks.

**Supporting References**

[1] M. D. Tran, J. H. Kim, H. Kim, M. H. Doan, D. L. Duong, Y. H. Lee, *ACS Appl. Mater. Interfaces* **2018**, *10*, 10580.

[2] H. Wang, C. Zhang, F. Rana, *Nano Lett.* **2015**, *15*, 339.

[3] L. Yuan, L. Huang, *Nanoscale* **2015**, *7*, 7402.